\documentclass[a4paper,11pt]{article}
 \usepackage[a4paper,portrait,margin=1in]{geometry}
 \usepackage[colorlinks=true,allcolors=black]{hyperref}
 \usepackage[usenames,dvipsnames]{xcolor}
 \usepackage{graphicx} 
 \usepackage{amsmath,amssymb,mathrsfs}
 \usepackage{natbib}

  \newcommand{\ncra}{\href{http://www.ncra.tifr.res.in/}{National Centre for Radio Astrophysics (NCRA-TIFR)}}

  \newcommand{\cms}{\href{http://cms.unipune.ac.in/}{Centre for Modeling and Simulation}}
  \newcommand{\sppustat}{\href{http://stats.unipune.ac.in/}{Department of Statistics}}
  \newcommand{\sppu}{\href{http://www.unipune.ac.in/}{Savitribai Phule Pune University}}
  \newcommand{\sppuaddress}{Ganesh Khind, Pune 411007 India}

  \newcommand{\imsc}{\href{http://www.imsc.res.in/}{The Institute of Mathematical Sciences, HBNI}}
  \newcommand{\imscaddress}{Taramani, Chennai 600113 India}

  \newcommand{\mihir}{\href{http://cms.unipune.ac.in/~mihir}{Mihir Arjunwadkar}}
  \newcommand{\akanksha}{\href{http://stats.unipune.ac.in/images/pdf/ASK.pdf}{Akanksha Kashikar}}
  \newcommand{\manjari}{\href{http://www.imsc.res.in/~manjari/}{Manjari Bagchi}}

  \newcommand{\sushan}{\href{http://ncra.tifr.res.in/ncra/people/academic/visiting-scientists/Sushan_Konar}{Sushan Konar}}
  \newcommand{\dipanjan}{\href{http://ncra.tifr.res.in/ncra/people/academic/ncra-faculty/dmitra}{Dipanjan Mitra}}

\begin{document}

\title{Neutron stars in the light of SKA:\\ Data, statistics, and science}
\author{\mihir\\{\small \cms}\\{\small \sppu}\\{\small \sppuaddress}\\{\small \texttt{mihir@cms.unipune.ac.in}}\\\\\akanksha\\{\small \sppustat}\\{\small \sppu}\\{\small \sppuaddress}\\{\small \texttt{akanksha.kashikar@gmail.com}}\\\\\manjari\\{\small \imsc}\\{\small \imscaddress}\\{\small \texttt{manjari@imsc.res.in}}}
\maketitle

\begin{abstract}
The Square Kilometre Array (SKA), when it becomes functional, is expected to enrich neutron star (NS) catalogues by at least an order of magnitude over their current state.
This includes the discovery of new NS objects leading to better sampling of under-represented NS categories, precision measurements of intrinsic properties such as spin period and magnetic field, as also data on related phenomena such as microstructure, nulling, glitching, etc.
This will present a unique opportunity to seek answers to interesting and fundamental questions about the extreme physics underlying these exotic objects in the universe.
In this paper, we first present a meta-analysis (from a methodological viewpoint) of statistical analyses performed using existing NS data, with a two-fold goal:
First, this should bring out how statistical models and methods are shaped and dictated by the science problem being addressed.
Second, it is hoped that these analyses will provide useful starting points for deeper analyses involving richer data from SKA whenever it becomes available.
We also describe a few other areas of NS science which we believe will benefit from SKA which are of interest to the Indian NS community.

\vspace{1.0em} \noindent

\vspace{1.0em} \noindent
{\sf \em \scriptsize
\begin{flushright}
 To appear in \href{http://www.ias.ac.in/jaa/}{Journal of Astrophysics and Astronomy (JOAA)}

 special issue on ``Science with the SKA: an Indian perspective''

\end{flushright}
}
\end{abstract}

\section{Introduction}
\label{s:intro}

The Square Kilometre Array (SKA; \url{https://www.skatelescope.org/}) is a giant radio telescope in the making,
with an effective collecting area of approximately one square kilometre and operational frequency range between about 70 MHz and 10 GHz.
The low-frequency part of the observatory will be built at a site in Western Australia, while the mid- and high-frequency parts will be built in Southern Africa.
Probing fundamental physics using pulsars is one of the eleven key science goals of SKA (\url{https://www.skatelescope.org/science/}).
Building SKA is a huge technological challenge, and utilising its full capacity will be a scientific challenge.
As such, substantial international effort is being put into exploring different aspects of SKA science using its precursors and pathfinders.
The upgraded Giant Metrewave Radio Telescope (uGMRT), located near Pune, India and operated by the \ncra, is one such SKA pathfinder.
These developments have inspired the Indian neutron star (NS) community to engage in both the endeavours, SKA as well as uGMRT. 

SKA is expected to advance our understanding of the physics of NS through extensive high-quality data.
For example, we expect discovery of many new radio-emitting NS objects (leading to better sampling of under-represented NS categories)
together with precision measurements of their intrinsic properties such as spin period and spin period derivative.
In fact, the number of such objects is expected to increase by at least an order of magnitude (\citet{keane15}).
Other areas of advancement include intrinsic red noise in timing spectra as well as correlated noise due to stochastic gravitational wave background,
variation of the dispersion measure over time, etc.
This, in turn, is expected to lead to better understanding about derived parameters such as dipolar magnetic field at the surface, characteristic age, etc.
We also expect better data on pulsar phenomena such as microstructure, nulling, mode-changing, and glitching.
All together, this will present a unique opportunity to seek answers to interesting and fundamental questions about the extreme physics underlying these exotic objects in the universe.

With data rates, sizes, and complexities soaring up high over the coming decades,
meaningful investigation into a scientific question will require fresh ways of identifying patterns and structure in the data using sophisticated statistical and computational methodologies.
(Indeed, the 21st century science has been aptly described as \emph{large datasets, complex questions} science \citep{Efron2011}.)
Technology development, which is essential for progress of science, also necessitates methodological development for its efficient and effective use.
It is important to remember that it is the nature of the data and scientific questions being addressed which should dictate the method, and not \emph{vice versa}.

The present article, in part, intends to showcase some of the work being carried out (Sec.\ \ref{s:taxonomy}--\ref{s:microstructure}) by us.
A few other areas of NS science of interest to the Indian NS community, which we believe will benefit from the SKA data deluge and advanced statistical methods, are described in Sec.\ \ref{s:manjari1}--\ref{s:manjari2}.
This is not intended to be an extensive review about the use of statistics in neutron star astrophysics:
Through the case studies presented here, we hope to convey the challenges involved in devising or adopting statistical methods in the light of the questions being investigated.

\section{Taxonomy of neutron stars}
\label{s:taxonomy}

What different categories of NS objects are suggested by the available data?
Data-driven taxonomy of NS objects is an imminent direction for methodological research in the light of SKA.
Such data-driven research can benefit immensely from the use of advanced statistical methods
for data modeling, identifying potential outliers,
clustering/classification together with investigation of the most typical or atypical elements in a class, density estimation, regression, and statistical/machine learning \citep{HTF2003,Wasserman2004,Wasserman2006}.

\begin{figure}
 \centering
 \includegraphics[width=\textwidth]{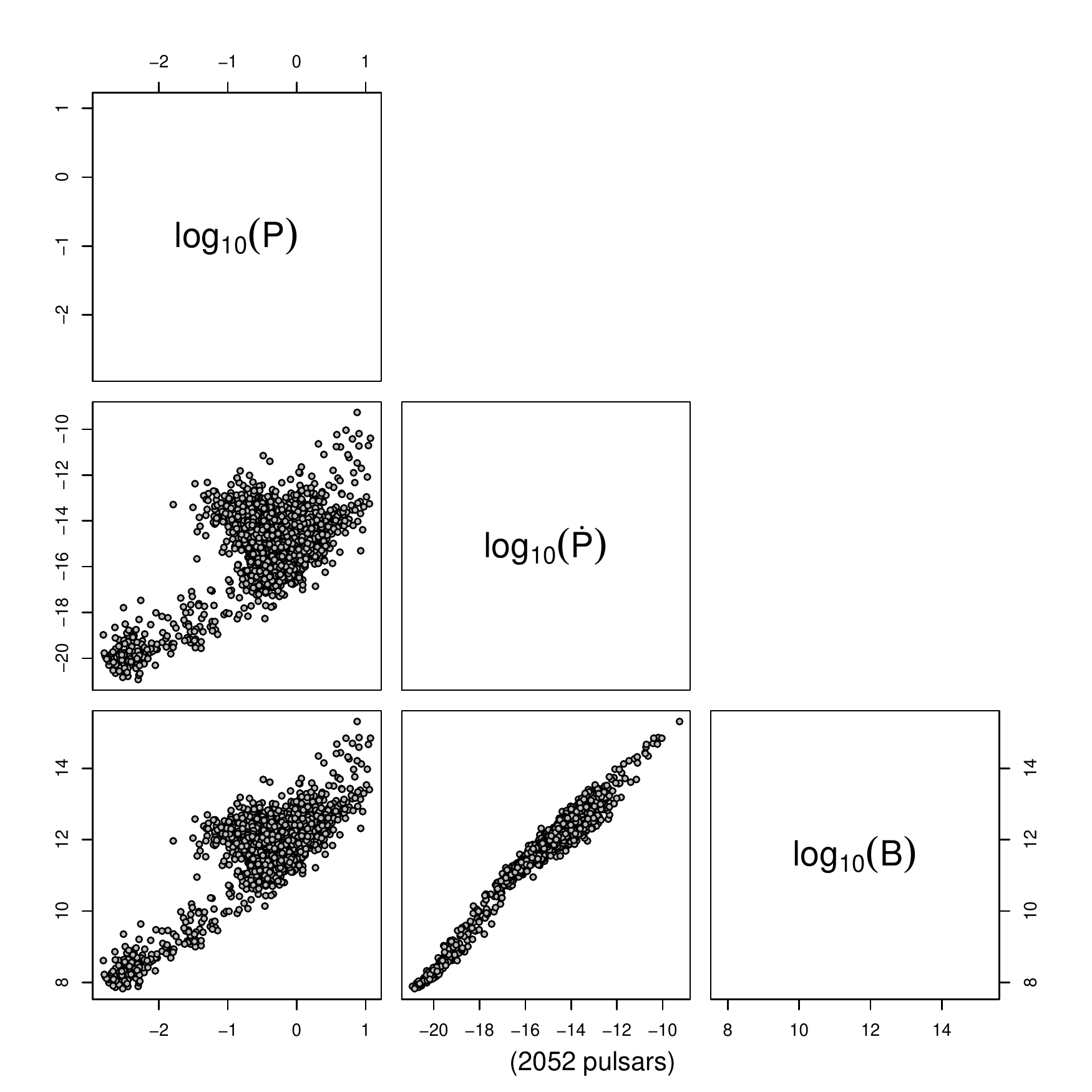}
 \caption{\label{f:logPlogB}Pairwise plots of three pulsar quantities: spin period ($P$ in second), spin-period derivative ($\dot{P}$), surface magnetic field ($B$ in Gauss). The overall look of the $P$--$\dot{P}$ (left middle) and $P$--$B$ (left bottom) plots is similar, which suggests that use of either pair of variables in cluster analysis will lead to similar results. Data source: The ATNF catalogue (\citet{MHT+2005}; \url{http://www.atnf.csiro.au/people/pulsar/psrcat/}; version 1.54, retrieved August 2016). See Sec.\ \ref{s:taxonomy} for discussion. Pulsars with undetermined (444) or non-positive (40) spin-period derivative $\dot{P}$ are not included in these plots.}
\end{figure}

From the NS theory point-of-view, it may be argued that two of the fundamental \emph{physical} properties of NS for which precise data are available,
namely, spin period $P$ and
the minimum possible dipolar magnetic field $B$ at the surface
(Fig.\ \ref{f:logPlogB}),
would be appropriate at the most basic level of such an analysis,
although spin period $P$ and period derivative $\dot{P}$ are the most fundamental \emph{measured} properties.
While we used $P$ and $B$ for the exercise described in this section,
any other appropriate set of measured or derived properties can be used for this purpose in principle and in practice.
As $B \propto \sqrt{P \dot{P}}$, and there is a weak correlation between $P$ and $\dot{P}$, most of the conclusions and drawbacks mentioned in the subsequent discussion of classification using $P$ and $B$ remain valid for classification carried out using $\dot{P}$ instead of $B$.
(We note that although \emph{precise}, any data involving $B$ may not be \emph{accurate}, because the assumptions behind the above relationship between $P, \dot{P}$ and $B$ will not always be valid.)

A work in progress \citep{KKA2016} explores machine learning methods to categorize NS based on the two fundamental intrinsic properties $P$ and $B$.
Specifically, we have explored a few standard clustering methods such as hierarchical clustering, $k$-means, $k$-medoids, etc.
Optimal number of clusters needs to be decided up-front in these approaches,
and most methods for this purpose rely on different criteria that take into account the within-cluster and between-cluster variability of clustering \citep{HTF2003}.
For available NS data,
the optimal number of clusters turns out to be 2 or 3, primarily separating regular pulsars from millisecond pulsars.
However, the within-cluster variability is quite high.
Further, because the two intrinsic properties are positively correlated (because $B \propto \sqrt{P\dot{P}}$),
it turns out that clustering carried out using magnetic field \emph{alone} provides nearly the same results.
This suggests that additional information on other variables, possibly orthogonal to the two existing ones, should provide better classification.

\begin{figure}
 \centering
 \includegraphics[width=\textwidth]{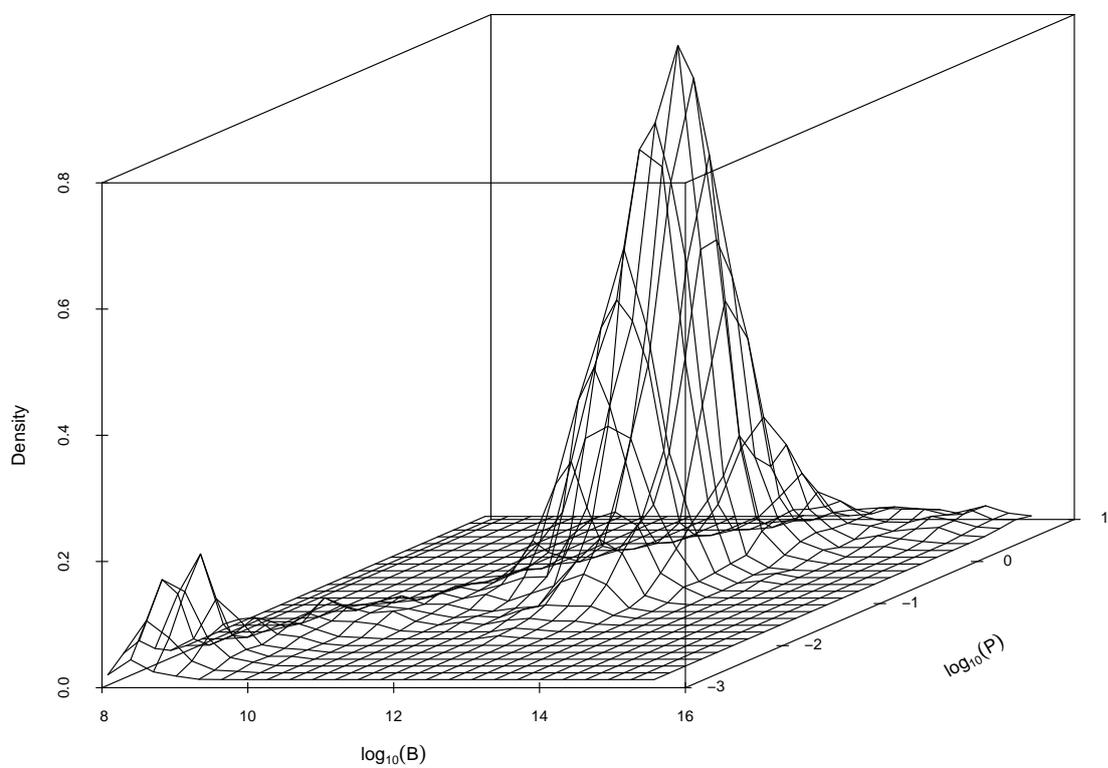}
 \caption{\label{f:BVDen}A wireframe plot of the estimated bivariate kernel density in the spin period ($P$ in second) -- surface magnetic field ($B$ in Gauss) plane. See Sec.\ \ref{s:taxonomy} for discussion.}
\end{figure}

The estimated probability density in the $\log_{10}(P)$--$\log_{10}(B)$ plane is shown in Fig.\ \ref{f:BVDen} in the form of a bivariate kernel density plot.
It shows two clear peaks corresponding to the two known broad groups (millisecond and non-millisecond pulsars).
Finer structure in the bivariate density function may be deciphered using advanced statistical methods such as
model-based clustering \citep{FR-MBC-2002}.
This may also result in the use of the correlation structure for enhancing the classification.

Supervised learning (i.e., classification) methods can be employed to assess the role of these variables in corroborating the existing (physics-based) classification. Our preliminary results (not shown) indicate that the existing NS groups are not linearly separable on the basis of these two variables alone: This again underscores the need for reliable data on other NS characteristics so as to improve data-driven classification.

Another approach to tackle this problem of overlapping groups is to go for soft-clustering methods.
The traditional cluster analysis algorithms create mutually exclusive clusters by assigning every observation to exactly one cluster.
However, in some situations, assigning an observation to more than one class might be more beneficial.
To see this, consider the following illustration.
While organizing documents on a computer, we use  mutually exclusive folders.
In contrast to this, labelling facility in a Gmail account allows you to give multiple labels to every email.
This is more useful as you might want to categorize an email into multiple categories.
For example, an email invoice for a flight booking might go into multiple categories such as Work, Financial, Travel etc.
Mixed membership models allow such soft clustering by assuming that individuals or observational units may only partly belong to population mixture categories.
These categories are referred to as topics, extreme profiles, pure or ideal types, states, or subpopulations, depending on the field of application.
The degree of membership is a vector of continuous non-negative latent variables that add up to 1 \citep{IntroMMM}.
The origin of this idea is thought to be in the Grade of Membership (GoM) model which allowed for fuzzy classifications in medical diagnosis problems \citep{WCG1978}.
In the context of NS, this will better help locate the boundary members for the clusters obtained.

Unequal class sizes pose yet another challenge with these data.
The performance of many supervised classification algorithms is adversely affected by variation in group sizes.
Further, commonly used measures of misclassification such as accuracy fail to take into account the effect of unequal group sizes.
For example, if 5 observations from a group of size 20 are misclassified, that results into a misclassification rate of $25\%$ for that particular group.
However, if the total number of observations is large, the overall misclassification rate is reported to be low due to less misclassification in the bigger groups.
Hence, while developing or adopting classification methods, this issue needs to be taken into consideration.

\section{Pulsar glitches: Multiple populations?}
\label{s:glitch}

Do the pulsar glitch data support the above view that there could be more than one mechanism responsible for producing glitches?
What do the data say about the glitch distribution having more than one mode?
These questions are addressed in \citet{KA2015-2} (see also \citet{KA2014}).
Because the estimated errorbars on glitch magnitudes are at least an order-of-magnitude smaller,
we decided to ignore them and to consider the glitch magnitudes as well-determined, precise data.
This can be thought of as a data-modeling assumption that simplifies the subsequent analysis,
because if the glitch data were not so precise, then what may appear multimodal in one dataset may not appear so in another.

Given the nature of the question, it was natural to consider formulating this problem as a statistical inference problem involving hypothesis tests
where the null hypothesis is that of unimodality.
The specific tests we applied to a recent version of a glitch dataset \citep{ELS+2011} were the dip test \citep{HH1985},
the Silverman test \citep{Silverman1981,HY2001},
and the bimodality test \citep{HV2008}.
The individual $p$-values of all tests suggest strong evidence against the null hypothesis of unimodality.
While applying a battery of hypothesis tests to a dataset may not seem like a good statistical practice,
this was done here for exploratory purposes only to arrive at tentative qualitative conclusions.

\begin{figure}
 \centering
 \includegraphics[width=1.05\textwidth]{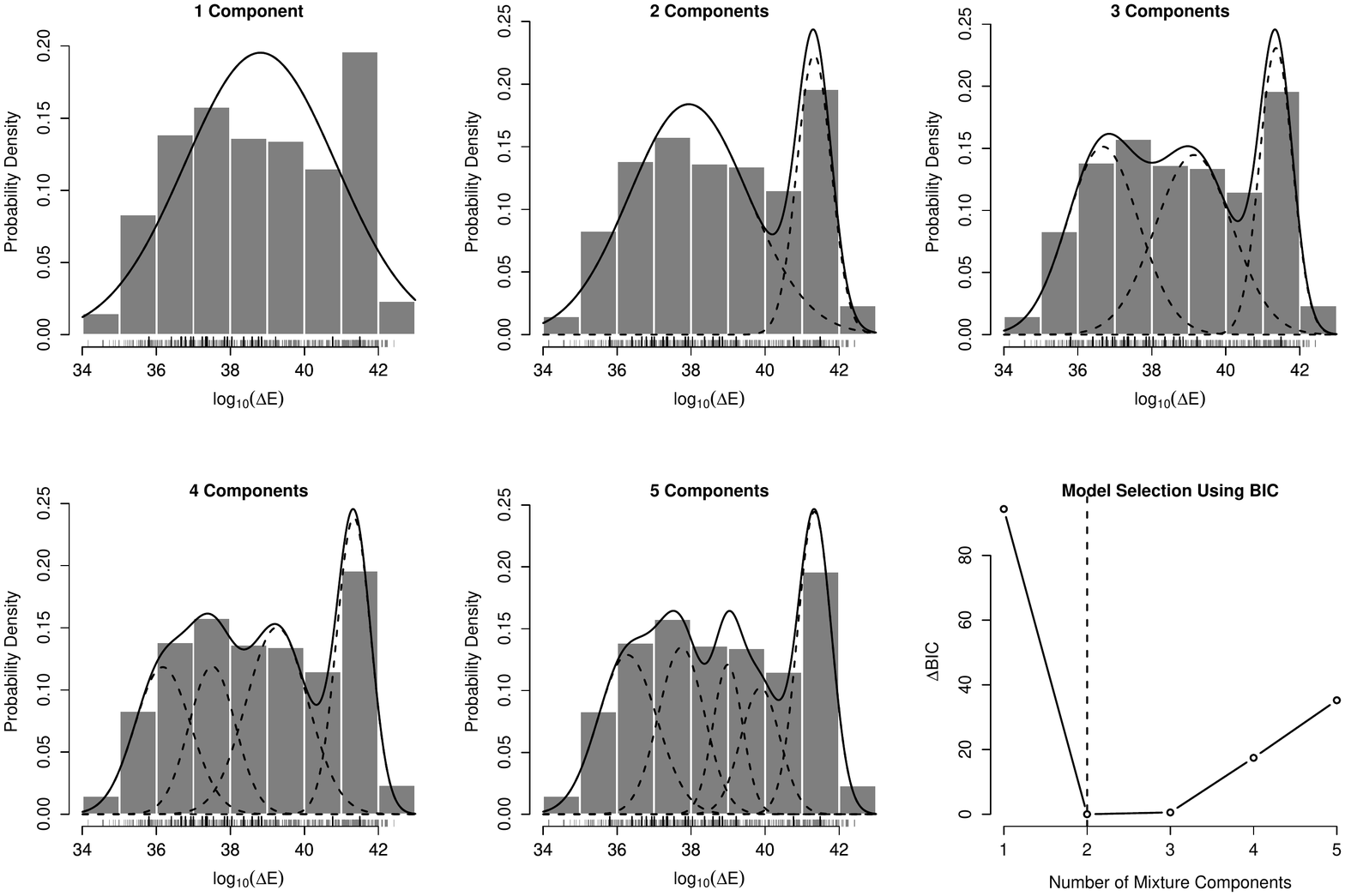}
 \caption{\label{f:glitch}Gaussian mixture fits (solid curves), together with BIC-based model selection (bottom right), to a pulsar glitch magnitude ($\Delta E \equiv I \nu^2 \times \delta \nu / \nu$) dataset (\citet{ELS+2011}; \url{http://www.jb.man.ac.uk/pulsar/glitches.html}; retrieved March 2015). Dashed curves represent individual Gaussian components of the fitted mixture. The BIC-optimal Gaussian mixture has 2 components for this dataset. Figure courtesy: \citet{KA2015-2}. See Sec.\ \ref{s:glitch} for discussion.}
\end{figure}

We also considered a more general scenario in which the data may be unimodal but has the structure suggestive of a mixture of multiple populations.
This suggested modeling the glitch data using a mixture model; specifically, a mixture of Gaussian probability distributions \citep{MP2000} each representing one glitch mechanism.
The Gaussianity assumption can be contested.
For example, self-organized criticality considerations (see, e.g., \citet{Aschwanden2011}), if applicable in the glitch context, may suggest a mixture of Pareto-like power-law distributions.
It was used here only to see if there is enough structure to glitch data to support the multiple-populations view.
Use of Gaussian mixtures, in particular, can lead to serious overfitting and identifiability problems, which could invalidate an analysis or render it inconclusive.
This pathology was avoided through judicious use of BIC-based model selection \citep{BA2002}.
Our results suggest structure in the glitch probability distribution which supports the multiple populations view (see Fig.\ \ref{f:glitch}).

\section{Pulsar null fraction}
\label{s:nulling}

Certain pulsars cease to emit pulses abruptly for one or more consecutive periods.
This phenomenon is called \emph{nulling} \citep{Backer1970}.
The proportion of pulses not emitted, which is a gross measure of the extent of nulling, is called the \emph{null fraction} (NF).
The phenomenon of nulling, along with related phenomena such as mode-changing and intermittency,
are important for an understanding of the emission mechanisms operative in pulsars.
The portions of a pulse with and without appreciable emission are called, respectively, the \emph{on-pulse} and \emph{off-pulse} parts.
NF is conventionally estimated using pulse energy data, by matching off-pulse and on-pulse energy histograms \citep{Ritchings1976}.
This approach, although simple and popular, relies on histograms and thereby on an \emph{ad hoc} choice of the histogram bin width.
Such \emph{ad hoc} approaches that do not have firm grounding in statistical theory come with little or no formal guaranties
(e.g., regarding their behaviour in the asymptotic regime, contamination in the data, etc.).
Further, obtaining errorbars in the form of valid confidence intervals is usually difficult in such approaches.

\begin{figure}
 \centering
 \includegraphics[width=\textwidth]{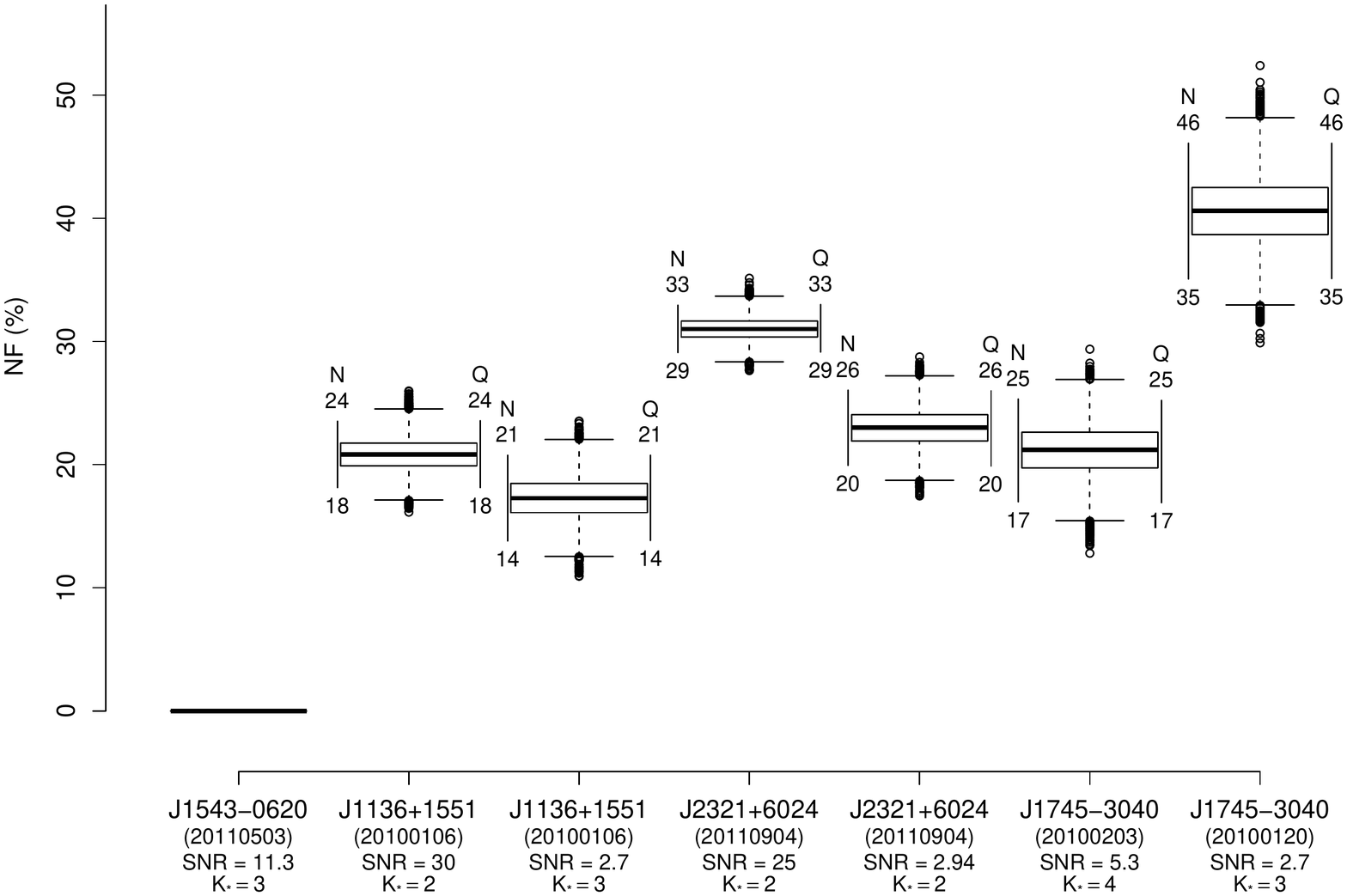}
 \caption{\label{f:nf}95\% confidence intervals on the null fraction for a set of well-characterized regular (i.e., non-millisecond) pulsars, obtained via bootstrap based on BIC-optimal Gaussian mixture fits to archival datasets from the GMRT \citep{SAK+1991}. Boxplots represent bootstrap distributions for NF. Vertical lines on either side of a boxplot represent two kinds of bootstrap confidence interval (N $\equiv$ normal-based, Q $\equiv$ quantile). $K_*$ in the $x$-axis labels indicates the BIC-optimal Gaussian mixture size, while the parenthesized date indicates the epoch of observation. The confidence intervals for all datasets (except J1745-3040) are consistent with the literature values (not shown here; see \citet{ARG2016} for further details). These results serve to validate the methodology used, as well as to show its usefulness even for low-S/N datasets. See Sec.\ \ref{s:nulling} for discussion. For the J1543-0620 data set, all the bootstrap NF estimates turn out to be \emph{exactly} zero; the confidence interval on NF is, therefore, [0,0].}
\end{figure}

This problem of estimating NF together with valid confidence intervals is considered in \citet{ARG2016},
where we use Gaussian mixture models \citep{MP2000} to model the on-pulse energy distribution.
This methodology comprises of Gaussian mixtures to model the pulse energy data,
a robust multivariate method to identify outliers in the data,
the EM algorithm to fit models to the data,
BIC-based model selection \citep{BA2002} to choose an optimal mixture model,
and two bootstrap prescriptions for computing confidence intervals on the null fraction \citep{Wasserman2004}.
We have presented results on
archival Giant Metrewave Radio Telescope (GMRT) \citep{SAK+1991} data for a set of well-characterized pulsars to illustrate and validate the methodology (Fig.\ \ref{f:nf}).

A key feature of this methodology is the use of off-pulse energy distribution parameters to constrain the null-pulse Gaussian in the mixture model.
Together with BIC-based model selection, this use of available information helps alleviate the identifiability and overfitting problems notoriously associated with Gaussian mixtures.
Compared to the conventional method, this methodology works well even for data with moderately low signal-to-noise ratio (S/N).
Indeed, a companion paper \citep{RAGK2016} applies this methodology to investigate nulling in millisecond pulsars where the S/N is low.

This methodology can be criticized on two counts.
First, the computational cost, which scales with the data size and the number of bootstrap replications, can be high especially for large datasets.
(Thankfully, the bootstrap is a happily parallel computation.)
Second, from a statistical modeling viewpoint, is such detailed model as a Gaussian mixture really necessary if the purpose is only to obtain NF?
Perhaps not.
Our hope is that, in future, such detailed modeling of the on-pulse energy distribution, which can be thought of as a probabilistic description of the emission profile of a pulsar, might turn out to be useful as device for characterizing or classifying pulsars.

\section{Pulse microstructure}
\label{s:microstructure}

Sufficiently bright single pulses from many pulsars, when observed with sufficient time resolution, show intensity fluctuations over longer
timescales of the order of the pulse period, and shorter timescales of the order of tens of microseconds.
The latter, which are quasiperiodic intensity variations, are called the \emph{microstructure} of a pulse \citep{CCD1968}.
Individual single pulses show considerable variability over and above the average pulse profile on both timescales.
The microstructure timescale is characteristic to a pulsar.
Both timescales (microstructure and pulsar period) and their relationship to each other are important from the perspective of pulsar emissions mechanisms.

\begin{figure}
 \centering
 \includegraphics[width=0.88\textwidth]{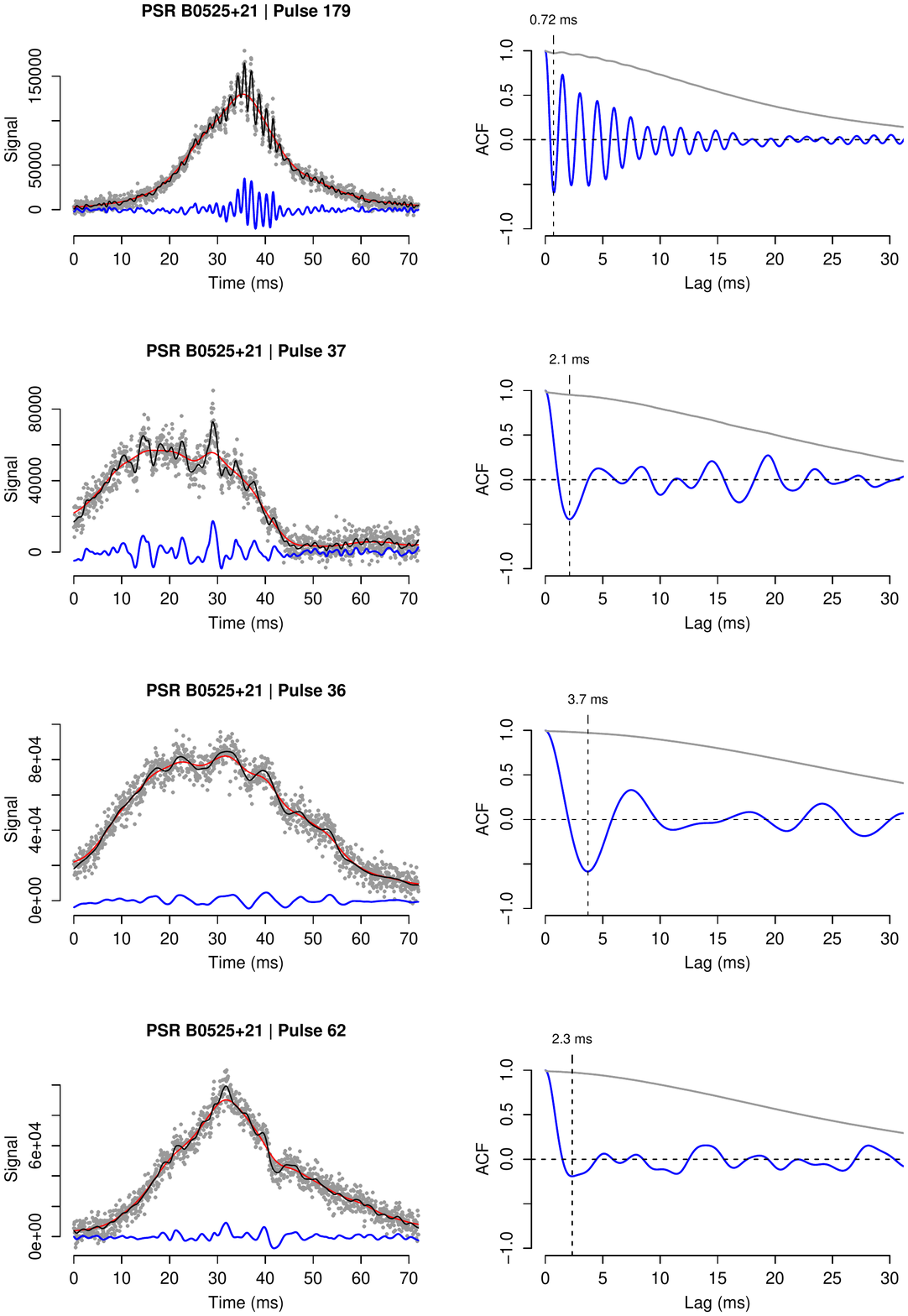}
 \caption{\label{f:microstructure}Estimating subpulse microstructure timescales. Figure shows ``a few illustrative subpulses from a B0525+21 dataset and their respective ACFs. Left column: gray, subpulse time series data; black, probable signal in the data sans the noise; red, probable smooth trend in the time series; blue, probable microstructure feature of the signal. Right column: gray, ACF computed directly from noisy data; blue: ACF computed from the probable microstructure feature''. First minimum in the ACF represents a timescale for the quasiperiodic microstructure oscillations. This minimum is much better resolved when microstructure is isolated from smooth trend and noise in the subpulse data, which is what the methodology proposed in \citep{MAR2015} attempts to do. Figure and part of the caption are reproduced here with permission from the authors of \citet{MAR2015}. See Sec.\ \ref{s:microstructure} for discussion.}
\end{figure}

The problem of estimating the microstructure timescale is considered in \citet{MAR2015}.
The traditional device for extracting microstructure timescales is the autocorrelation function (ACF),
where timescales are estimated as minima or maxima in the ACF (Fig.\ \ref{f:microstructure}).
The procedure is confounded by the fact that the ACF of a pulse is dominated by power at low frequencies;
i.e., by the longer-timescale variations, which we refer to as the \emph{envelope}.
In this frequency-domain view, microstructure corresponds to higher-frequency but low-power features in the power spectrum.
Additional oscillations can result from the presence of noise in the pulse time series.
To be able to estimate the microstructure timescale using ACF, one therefore needs to isolate the microstructure component
from the noise and the envelope components.
After considering many different methods both in the time and frequency domains, the strategy which worked best was as follows:
First, obtain a model-independent smoothing spline fit with optimal smoothing; this gives a denoised version of the pulse.
Second, estimate the envelope using kernel regression with a heuristic bandwidth which smoothens out the microstructure component completely.
The difference in these two fits is taken to be the microstructure component, which is used to estimate timescales via the ACF approach.

The strength of this approach is that it tries to neatly separate noise, envelope, and microstructure components of a pulse time series.
Shortcomings of this approach are (a) that the envelope is obtained using a heuristic bandwidth obtained through trial-and-error; and
(b) that formal analysis of the behaviour of our complete prescription has not been done.
Despite this, the approach is turning out to be useful in analyzing single pulses (see, e.g., \citet{MRA2016}),
and we hope that its limits of applicability as well as better alternatives will emerge with its continued use by the pulsar community.

\section{Pulsar spectral indices and luminosity distribution}
\label{s:manjari1}

\begin{figure}
 \centering
 \includegraphics[width=\textwidth]{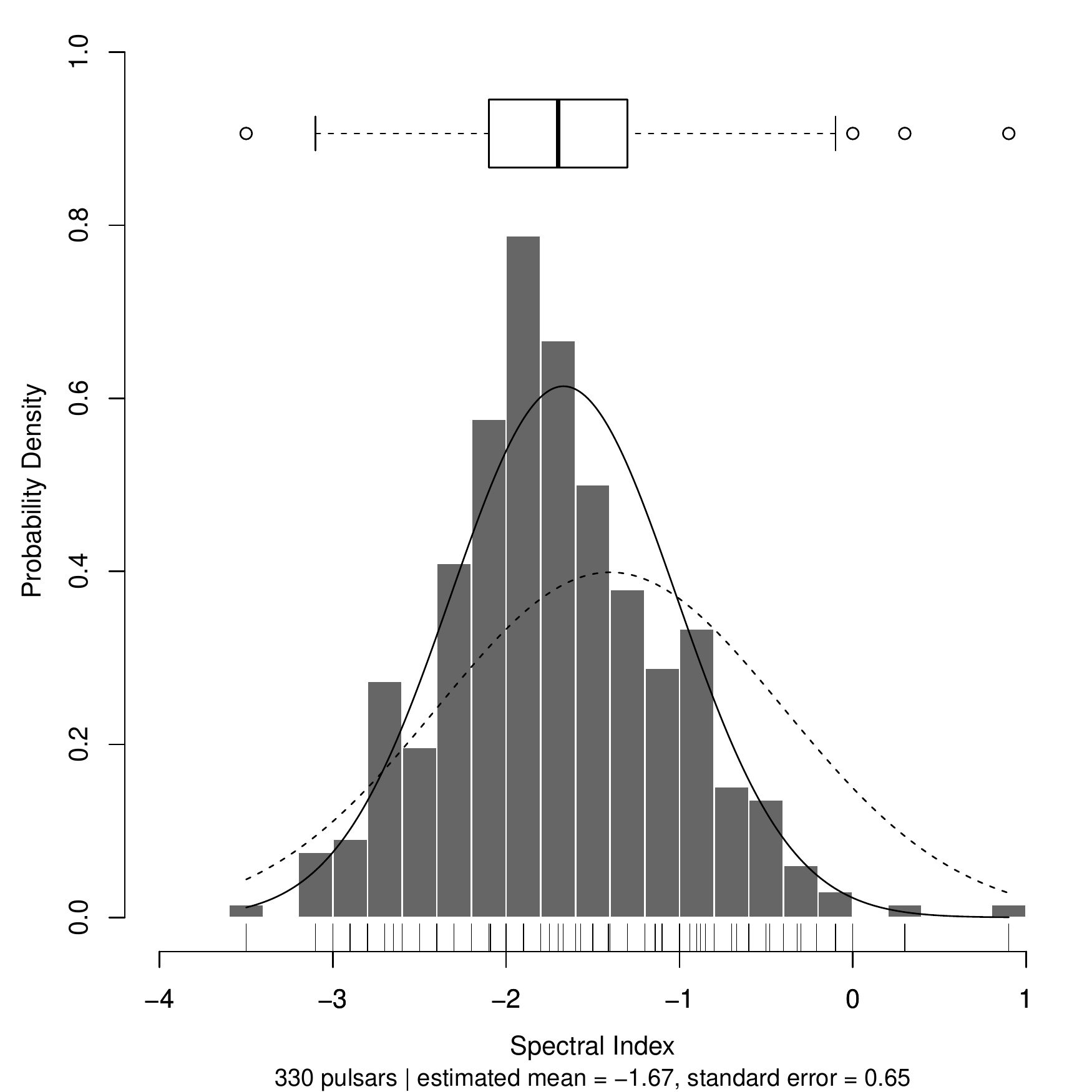}
 \caption{\label{f:alpha}Distribution of spectral indices available in the ATNF catalogue (\citet{MHT+2005}; \url{http://www.atnf.csiro.au/people/pulsar/psrcat/}; version 1.54, retrieved May 2016). Solid curve: Gaussian probability density function with the same mean and standard error as that of this spectral index dataset; dashed curve: Gaussian probability density function with mean -1.4 and unit standard deviation as suggested in \citet{blv13}. See Sec.\ \ref{s:manjari1} for discussion. Overlaid on the histogram is a boxplot representation of the same data, indicating potential outliers as circles and the 0, 25, 50, 75, 100 percentiles (excluding the potential outliers) as vertical lines.}
\end{figure}

There has been significant improvement in the knowledge about pulsar spectral indices and luminosity distributions in recent years mostly through careful statistical analysis of available data.
Pulsars are not equally bright at all frequencies, and the frequency dependence of pulsar brightness is expressed as $S_{\nu} \propto \nu^{\alpha}$, where $S_{\nu}$ is the flux density of the pulsar at the observing frequency $\nu$ and $\alpha$ is known as the \emph{spectral index}. The value of $\alpha$ is different for different pulsars. For most pulsars, it lies in the range $-2$ to $-1$, implying that pulsars are in general brighter at lower frequencies.
Fig.\ \ref{f:alpha} shows the histogram of spectral indices for 330 pulsars, with estimated mean $=-1.67$ and estimated median $=-1.70$.
This dataset has extreme examples such as PSR
J0711+0931 ($\alpha=-3.5$) and PSR J1740+1000 ($\alpha= 0.9$). There are also a few pulsars with almost flat spectra with $\alpha$ between $+0.30$ to $-0.30$. Some pulsars even show deviations from a single power law; i.e., their luminosities peak as functions of frequency: For some pulsars, this peak is around 1 GHz \citep{kgk07}, while for some others the peak is seen around 100 MHz \citep{mgj94}. The subset of 39 millisecond pulsars (MSP) has less diverse values for the spectral index (from $-2.9$ to $-1.1$, with mean $-1.86$ and median $-1.80$). It is therefore conventional to use $\alpha=-1.9$ for MSPs \citep{tbms98, blc11}.
However, not many extensive statistical analyses of the distribution of $\alpha$ have been reported, mostly because the dataset is not large enough. One exception is the recent study by \citet{blv13}, who used a population synthesis technique in conjunction with likelihood-based analysis, and concluded that the distribution of $\alpha$ from different pulsar surveys can be modelled as a Gaussian distribution with mean $-1.4$ and unit standard deviation (see Fig.\ \ref{f:alpha} for comparison). They concluded that high-frequency ($>2$ GHz) surveys preferentially select flatter-spectrum pulsars, and that the opposite is true for lower-frequency ($<1$ GHz) surveys, and hypothesized that many known pulsars which have been detected at high frequencies will have small, or positive, spectral indices. More data on pulsar spectral indices coming out of flux density measurements over a wide frequency range, followed by extensive statistical analysis will therefore be useful in understanding this diversity and any relation between the spectral index and other properties of pulsars, eventually shedding more light into the complex pulsar emission mechanisms. SKA and its pathfinders (including uGMRT) can contribute significantly in addressing this issue.

The question of (pseudo-)luminosities $L_{\nu} = S_{\nu} d^2$, where $d$ is the distance, similarly deserves deeper investigation.
Earlier, $L_{\nu}$ used to be considered a function of the spin period $P$ and spin period derivative $\dot{P}$
(see \citet{bag13} for a recent review of spin-dependent luminosity models).
Of late, people have started using spin-independent luminosity distribution functions, the most popular of which have power-law or broken power-law forms.
A pioneering work in this direction is \citet{lbd93}, who found that that the distribution of total galactic population of pulsars with $L_{400} > 10 ~{\rm mJy ~kpc^2}$ could be expressed as $N (\geq L_{400} ) = (7.34 \pm 1.06) \times 10^4 L_{400}^{-1}$, whereas the distribution of potentially observable pulsars with $L_{400} > 10 ~{\rm mJy ~kpc^2}$ could be expressed as $N (\geq L_{400}) = (1.31 \pm 0.17) \times 10^4 L_{400}^{-1}$.
In both cases, $N (\geq L_{400})$ is the total number of pulsars having luminosities equal to or greater than $L_{400}$.
Later studies \citep{fk06, blc11} which attempted to assess the agreement between simulated and observed data via the Kolmogorov-Smirnov test preferred the log-normal distribution with mean $-1.1$ and standard deviation $0.9$, although they could not rule out power-law models.

Clearly, this is yet another area where substantial progress can be expected through the use of advanced statistical models and methods on larger datasets. Awaiting SKA, the current state of understanding can be refined using data from ongoing and future pulsar surveys from different telescopes, such as the all-sky survey using LOFAR and a similar proposed survey using uGMRT.

\section{Bayesian statistics in Pulsar astronomy}
\label{s:manjari2}

Bayesian statistics \citep{wal03,gre05} is becoming increasingly popular in pulsar astronomy like in many other branches of astronomy.
In this section, we discuss a few recent advances in pulsar astronomy based on  Bayesian approaches.

Notable applications of Bayesian parameter estimation over the last few years include prediction of observable population of pulsars in Galactic globular clusters \citep{clmb13}, large Magellanic clouds \citep{rcl13}, and close to the Galactic centre \citep{cl14}, all using the presently known population as the prior. Another example is the prediction of the Galactic NS--NS merger rate using the data on double pulsars  as prior information \citep{kpm15}. Bayesian analysis has been employed to understand the uncertainties in pulsar distance measurements due to Lutz-Kelker bias \citep{vlm10, vwc12}. As a better prior improves Bayesian parameter estimates, it would be highly desirable to re-do all these analyses in the SKA era. Presently, we are exploring the properties of black-widow and red-back pulsars within the framework of Bayesian parameter estimation.

One notable application of Bayesian model selection followed by parameter estimation is in high precision pulsar timing analysis. The primary motivation here is the detection of nHZ gravitational waves.
Conventional pulsar timing models include effects such as pulsar spin down rate, delays resulting from the interstellar medium, motion of the earth, effects of the solar system planets, classical and general relativistic effects in case of binary motions, etc., which are incorporated in the standard pulsar timing package \texttt{TEMPO2}.
However, if additional noise processes such as red spin noise due to rotational irregularities, variation of dispersion delay with time or observing frequency, or correlated noise due to stochastic gravitational background, etc., exist, then those noises would lead to inaccurate estimates of timing parameters.
The complication here is that there is no way to know \emph{beforehand} whether any or some or all of these exist for a particular pulsar.
Bayesian model selection followed by parameter estimation would be useful in such cases, as shown by \citet{lent14}.
The analysis pipeline \texttt{TempoNest} created by them is freely available online (\url{https://github.com/LindleyLentati/TempoNest}).
Bayesian model selection followed by parameter estimation has been applied in studies of possible detection of gravitational waves of individual in-spiralling supermassive black-hole binaries using the timing data of an ensemble of pulsars, i.e., the pulsar timing array \citep[and references therein]{ellis13, teg14}.
The analysis pipeline \texttt{PAL2}, devised for this purpose is also publicly available (\url{https://github.com/jellis18/PAL2}).

Bayesian model selection is a complex set of methodologies that offers many possibilities. It would be desirable to further explore these thoroughly to assess their suitability for problems related to pulsar timing analysis.
The members of the \textit{Indian Pulsar Timing Array} (InPTA) group, including one of the authors (MB), intend to contribute to \texttt{PAL2} and \texttt{TempoNest}, as well as to explore further applications of Bayesian model selection plus parameter estimation in this area. We believe that this direction requires substantial development of analysis pipelines and software and, consequently, should provide ample opportunity for young aspiring researchers to contribute meaningfully.

\section{Conclusion}

At the risk of sounding pedagogic, the two most important lessons learnt during the course of the work discussed in this paper are as follows.
First, to understand their limits and suitability, methodologies -- old or new -- need to be explored and tested thoroughly on available data (if possible, from different observational instruments), while keeping in mind that with SKA on the horizon, eventually the data volumes may be much larger.
Second, involvement of researchers from allied fields (such as statistics and computer/computational science) who hold a genuinely multidisciplinary outlook and are open to stepping outside of their traditional comfort zones will help bring in fresh methodological perspectives.
This will go a long way in utilizing the full potential of the SKA for answering fundamental questions about the fundamental physics of neutron stars.

\section*{Acknowledgements}

MA would like to thank \sushan\ and \dipanjan\ for discussions, encouragement, and critical comments.
We would like to thank the anonymous referee for a thorough and meticulous review and for useful suggestions that have helped improve this paper.

\bibliographystyle{natbib}
\bibliography{ns-ska}

\newpage

\newpage

\newpage

\newpage

\newpage

\newpage

\end{document}